\documentclass{iaus}
\usepackage{graphicx}

\title[Kinematics and Composition] 
{Kinematics and Composition of the Galactic
Bulge: Recent Progress}

\author[Rich et al.]
{R. Michael Rich$^1$ 
\thanks{Supported by
NSF grant AST-0709479 and based on 
observations obtained at the W.M. Keck Observatory
and CTIO}, 
Christian Howard$^1$;
David B. Reitzel$^1$; HongSheng Zhao$^2$ \and Roberto de Propris $^3$}

\affiliation{$^1$Department of Physics and Astronomy,
UCLA, Los Angeles CA, 90095-1547 \break 
\break email: rmr@astro.ucla.edu; howard@astro.ucla.edu; reitzel@astro.ucla.edu\\
[\affilskip]
$^2$SUPA, School of Physics and Astronomy, University of St Andrews, KY16 9SS, UK
\break email: hz4@st-andrews.ac.uk\\[\affilskip]
$^3$ Cerro Tololo Inter-American Observatory, Casilla
603, La Serena, Chile 
\break email: rdepropris@ctio.noao.edu\\[\affilskip]
}

\pubyear{2008}
\volume{245}  
\pagerange{}
\date{?? and in revised form ??}
\jname{Formation and Evolution of Galaxy Bulges}
\editors{M. Bureau, E. Athanassoula, \& B. Barbuy, eds.}
\begin{document}

\maketitle

\begin{abstract}
We present recent results from
a Keck study of the composition of the
Galactic bulge, as well as results from
the bulge Bulge Radial Velocity Assay (BRAVA).
Culminating a 10 year investigation, Fulbright, McWilliam, \& Rich (2006, 2007) solved the problem of deriving the iron abundance in the Galactic bulge,
and find enhanced alpha element abundances, consistent
with the earlier work of McWilliam \& Rich (1994). 
We also report on a radial velocity survey of {\sl 2MASS}-selected M giant stars in the
Galactic bulge, observed with the  CTIO 4m Hydra multi-object spectrograph.  This program is to test dynamical models of the bulge and to search for and map any dynamically cold substructure in the
Galactic bulge.   We show initial results on 
fields at $-10^{\circ} < l <+10^{\circ}$ and $b=-4^{\circ}$.  We construct a
longitude-velocity plot for the bulge stars and the model data, and find
that contrary to previous studies, the bulge does not
rotate as a solid body; from $-5^{\circ}<l<+5^{\circ}$ the rotation curve 
has a slope of 
$\approx 100\  km\ s^{-1}$  and flattens considerably
at greater $l$ and reaches a maximum rotation
of $45\ {km\ s^{-1}}$ (heliocentric) or
$\sim 70\ {km\ s^{-1}}$ (Galactocentric).  This rotation is slower than that predicted by
the dynamical model of Zhao (1996).  

\keywords{Galaxy: bulge, abundances, kinematics and dynamics, formation galaxies: bulges}
\end{abstract}

The unique status of the bulge as a stellar population was not firmly established by Baade's discovery of RR Lyrae stars(already known in globular clusters) but rather
by the discovery of huge numbers of M giants.
These were cataloged using the 4m prime focus grism at Cerro Tololo (Blanco,
McCarthy, \& Blanco 1984).  Discovered by the thousands in the bulge but rare in globular clusters, the M giants would unlock much of the nature of the population and its link to distant galaxies (Frogel \& Whitford 1987).   Significant advances also include Whitford's (1978) demonstration that the integrated light of the bulge resembles that of ellipticals, the first abundance survey using K giants (Rich 1988) and the demonstration that the abundance distribution fits the simple one zone model of chemical evolution (Rich 1990).  The first high resolution study of bulge giants (McWilliam \& Rich 1994) showed elevated Mg and alpha elements.  This inspired numerous theoretical
papers (e.g. Matteucci et al. 1999) that constrain from
the elevated alphas, a rapid formation timescale of $<1$ Gyr for the bulge.   Recent efforts (Fig 1)confirm our high alphas (e.g.  McWilliam \& Rich 2004; Rich \& Origlia 2005; Cunha \& Smith 2006; Fulbright et al. 2007; Lecureur et al. 2007; Rich \& Origlia 2007).  Parenthetically, it is amusing that the [O/Fe] bulge study of Zoccali et al. (2007) rediscovers the rapid bulge formation timescale, but was also the subject of an ESO press release claiming the ``discovery'' of rapid bulge formation.

\begin{figure}
\begin{center}
\includegraphics[width=1.0\textwidth]{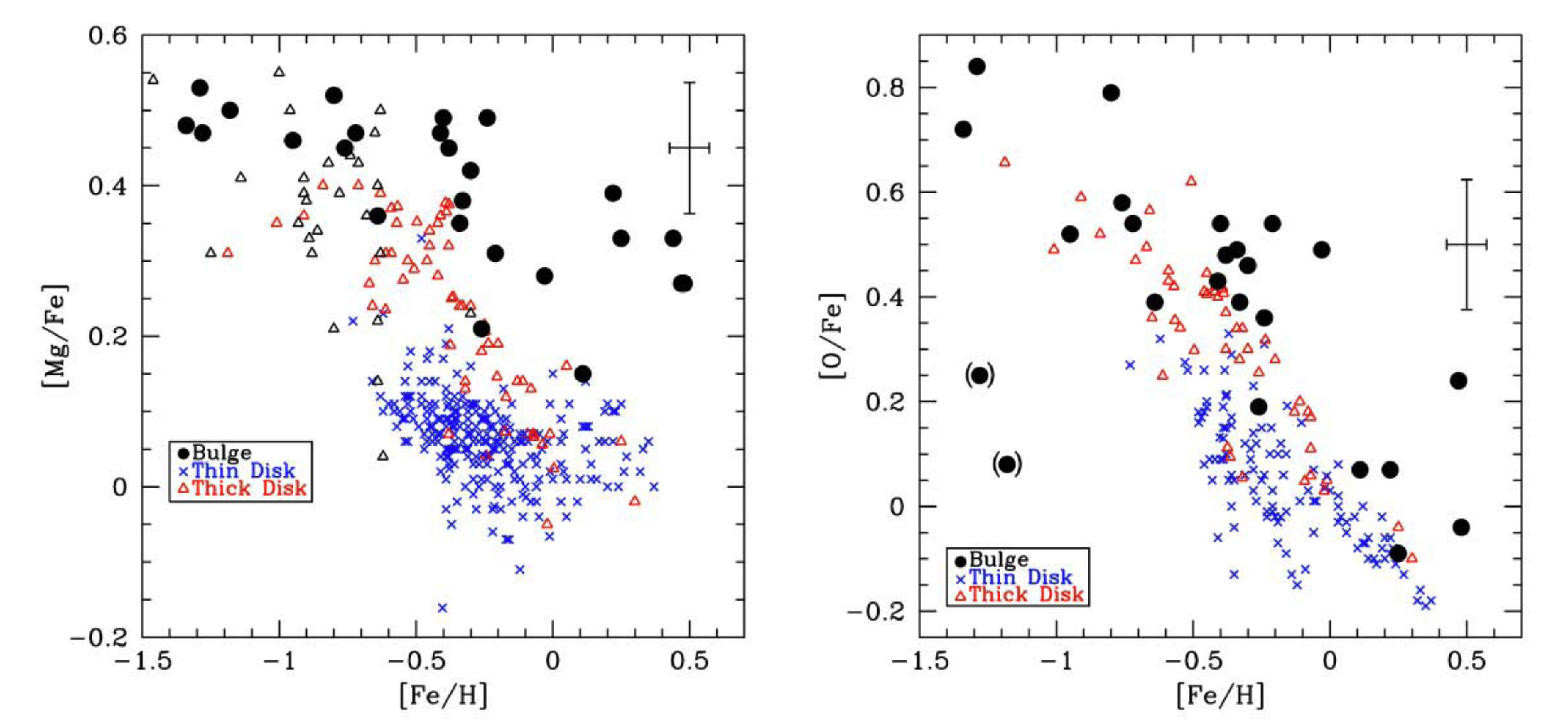}
\caption{{\it (Left):}  [Mg/Fe] vs [Fe/H] for
the bulge relative to thin and thick disk
populations (Fulbright et al. 2007).
This confirms a long established result (McWilliam \& Rich 1994) and is argued in chemical evolution models to support a 
rapid formation timescale of the bulge (cf. Matteucci et al. 1999) The same enhancement is also seen by Rich
\& Origlia (2005), Lecureur et al. (2007) and numerous
other studies.  {\it (Right):}
  Oxygen in the bulge, relative to thin and
  thick disk populations (see Fulbright et al. 2007 for
  details).    Oxygen is less  enhanced
  than Mg; this is unexpected since both elements should
  be synthesized in hydrostatic burning shells in massive stars.   McWilliam et al. 2007 propose that oxygen was never produced in the outer layers shed metal rich stars with $M>30M_\odot$  in Wolf-Rayet like mass loss.}
\end{center}
\end{figure}

\begin{figure}
\begin{center}
\includegraphics[width=1.0\textwidth]{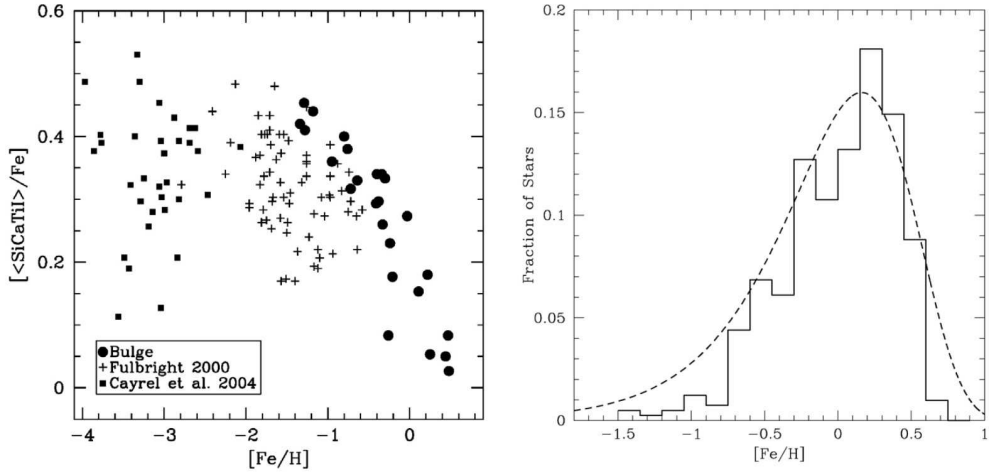}
\caption{{\it (Left):} Compared to two Galactic halo samples,
$\rm [<SiCaTi>/Fe]$ (alpha elements produced
in the SN explosion) are enhanced in the bulge.
Even at the lowest [Fe/H] in the bulge, the explosive
alphas define the upper envelope of enhancement 
relative to the halo (see Fulbright, McWilliam,
\& Rich 2007 for details).  {\it (right):} Fit of the Simple One Zone model of chemical evolution (Y=0.029) to an abundance
distribution of 409 bulge giants near $(l,b)=(0^\circ,-6^\circ)$; Minniti \& Zoccali 2007).  The result confirms the Rich (1990) finding that the Simple model is a good fit to the bulge abundance distribution.}
\end{center}
\end{figure}

\begin{figure}
\begin{center}
\includegraphics[height=5.0 cm]{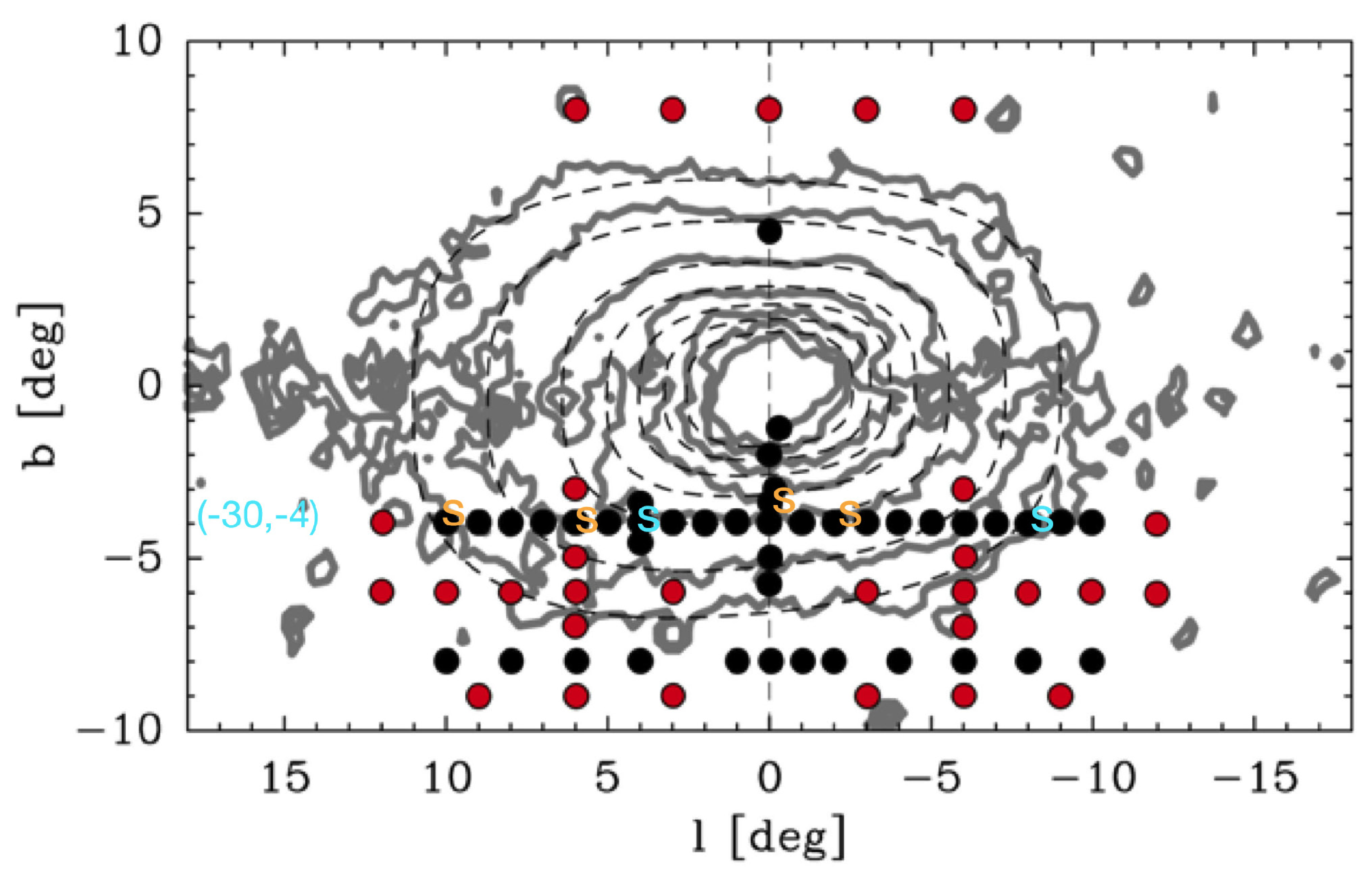}
  \caption{Galactic bulge fields in the {\sl BRAVA} survey, superposed on the bulge $2\mu \rm m$ map of
  Launhardt et al. 2002.  Points indicated in red are proposed for 2008 with
  aim of testing for cylindrical rotation and symmetry; `S' indicates $>2\sigma$ ``stream'' candidates.}
 \label{MB:Fig:BulgeColors}
\end{center}
\end{figure}

\begin{figure}
\begin{center}
\includegraphics[width=0.9\textwidth]{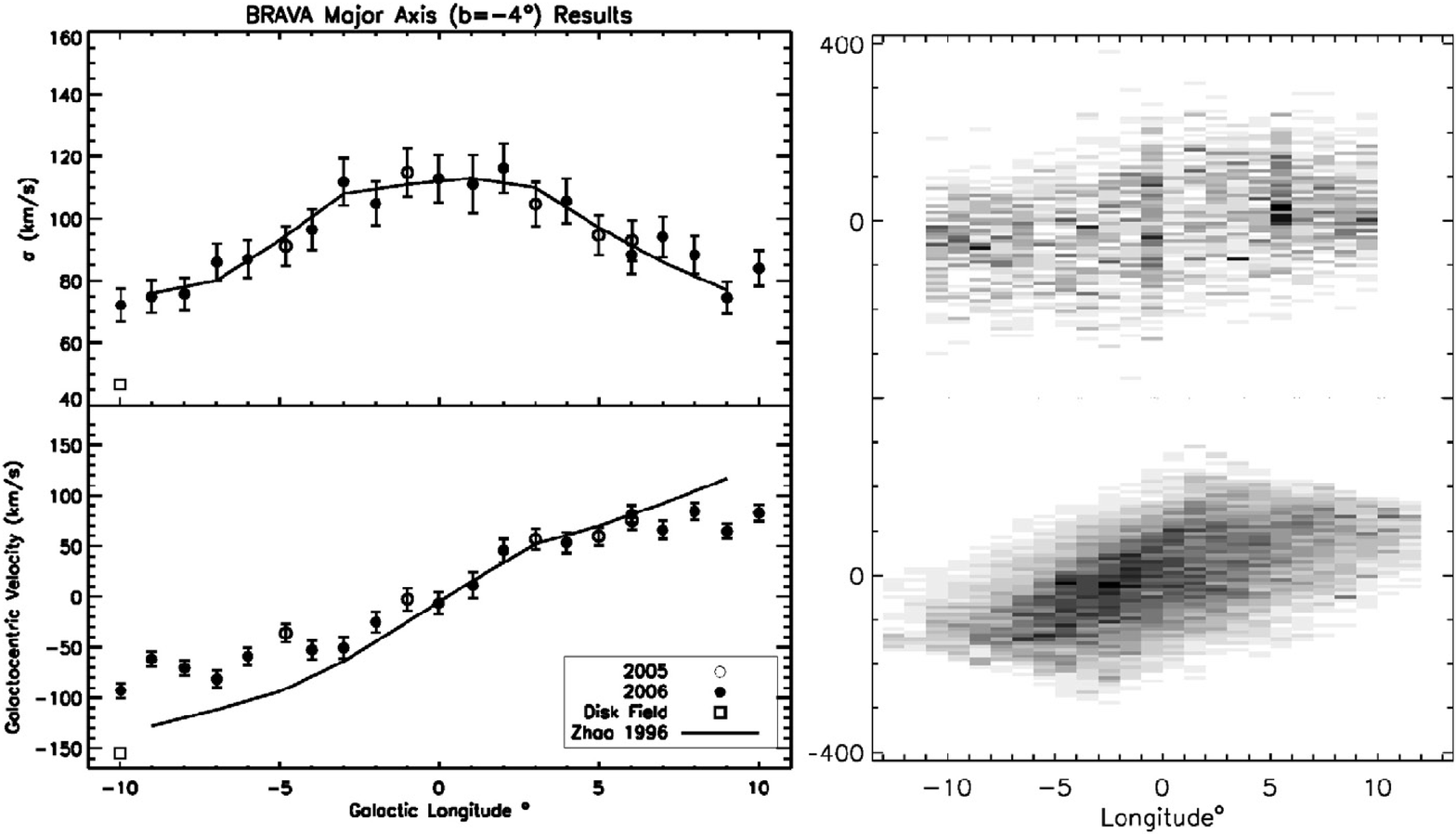}
  \caption{{\it (Left):} First and second moment results from
  {\sl BRAVA} including Galactocentric
  correction, compared to the Zhao (1996) model
  predictions.  Notice that no single solid body
  model fits this rotation curve.  
  $v,\sigma$ for the disk $(l,b)=-30^\circ,-4^\circ $ is indicated
  with an open square; disk contamination of {\sl BRAVA }
  is ruled out. {\it (Right):} 
  $l-v$ greyscale plot (heliocentric) for the bulge observations (upper)
  and the Zhao model (lower).  The slower rotation of the data
  is evident.  As the dataset improves the Zhao (1996) model will be adjusted to fit the data.}
\end{center}
\end{figure}

\begin{figure}
\begin{center}
\includegraphics[height=7.0cm]{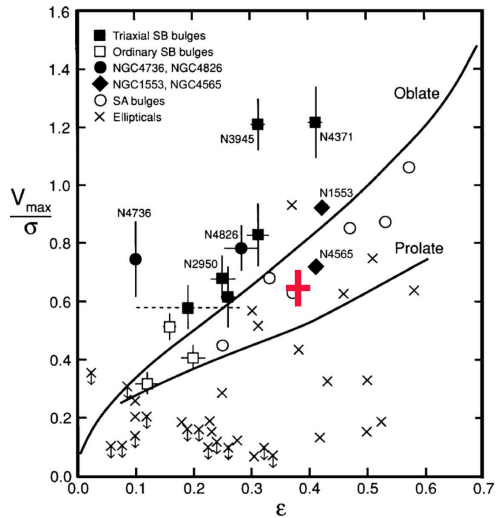}
  \caption{($V_{max}/\sigma$) plot from
  Kormendy \& Kennicutt (2004) with the
  Galactic bulge indicated (red cross).  The MW bulge lies under
  the oblate supported line and less rotationally supported
  than the pseudobulges (filled symbols) but is similar
  to classical bulges (open symbols).  }
\end{center}
\end{figure}

\begin{figure}[htbp]
\begin{center}
\includegraphics[height=7.0cm]{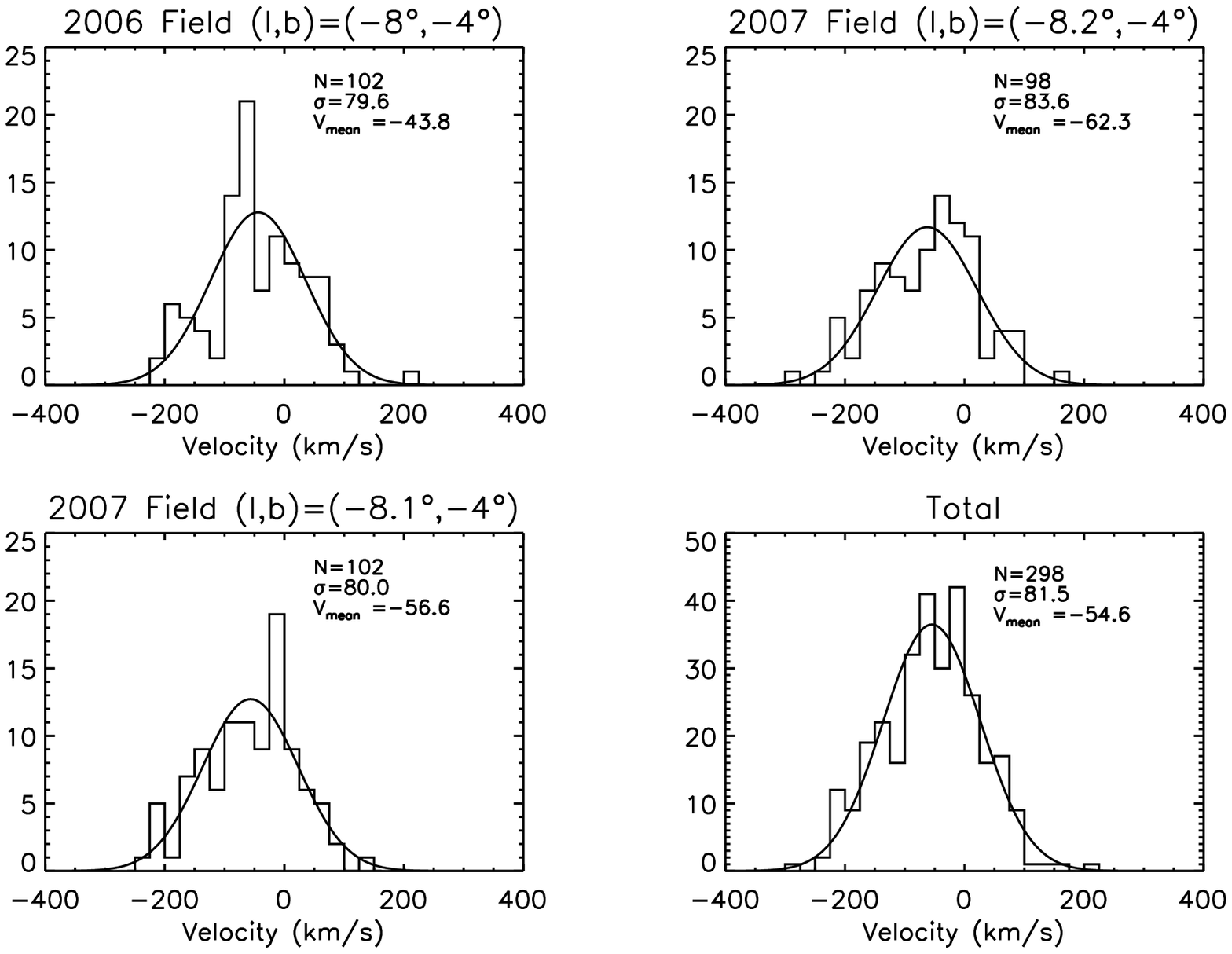}
\caption{A single season of observations (upper left)
finds a candidate 2.5 $\sigma$ cold kinematic feature
at $(l,b)=-8^\circ ,-4^\circ$.  New observations in 2007
did not confirm this feature; the summed velocity distribution appears Gaussian (lower right).  {\sl BRAVA} is finding a number of these candidates, each needing followup.}
\end{center}
\end{figure}

Recent optical abundance analyses were made possible by a novel solution of the iron abundance problem (Fulbright, McWilliam, \& Rich 2006) whereby the
extremely well studied red giant Arcturus takes the place of the Sun in providing physical oscillator strengths for iron lines that are weak enough that they retain their abundance sensitivity even at
high [Fe/H].   A differential abundance analysis between bulge giants and Arcturus effectively removes lingering concerns about evolutionary status, gravity, and most importantly, non-plane parallel atmospheres.   The proper solution of the iron abundance was a prerequisite to all subsequent optical studies of the composition.

The age of the bulge is hard to constrain, due to
reddening, spatial depth, contamination from
foreground disk stars, and an uncertain distance
modulus.   One option (Ortolani et al. 1995;
Zoccali et al. 2003) compares the bulge field age with
a metal rich globular cluster by force fitting the 
color-magnitude diagrams at the horizontal branch and main
sequence turnoff;
in the Zoccali study, the foreground disk was statistically
subtracted from the bulge yielding a globular cluster age
main sequence turnoff.  Kuijken \& Rich (2002) demonstrated that when the foreground disk is excised by a proper motion cut, the bulge shows a globular cluster-like turnoff and luminosity function.   In any case, young stars {\it are} present in the inner 100 pc and toward the nucleus
(cf. Figer et al. 2004).  At present, the rapid bulge formation timescale implied by the composition is consistent with constraints from the age.  Figure 2 also
shows that the recent high resolution survey of abundances (Minniti \& Zoccali 2007) is well fit by the
One Zone model (confirming Rich 1990); this is consistent with the rapid formation timescale mentioned earlier.

Figures 1 and 2, show some of our core results from
Fulbright, McWilliam, \& Rich (2007), including
the unexpected finding that oxygen is less enhanced
than Mg.   The alpha elements Si,
Ca, and Ti are thought to be produced in the SN explosion as opposed
to O and Mg that are produced in the hydrostatic shells before the SN.  The explosive alphas enhanced in the bulge relative to the halo over the full abundance range (Fig. 2).
Both Fulbright et al. (2007) and Lecureur et al. (2007)
note that Mg is enhanced even at high metallicity while
O shows a much less prominent enhancement,
following the disk trend.   Considering that both O and Mg are produced in the hydrostatic burning shells of massive stars, the result is of concern.  McWilliam \& Rich (2004) speculate that mass loss in the early
generation of massive stars via a Wolf-Rayet like
mechanism might be responsible.   Incorporating such
mass loss for massive metal rich stars, Maeder (1992)
find lower O yields; McWilliam et al. 2007 incorporate
these into new chemical evolution models that now
are consistent with the O vs Mg trends.  While further
confirmation is important, the O/Mg problem now has
an acceptable explanation.

There have been many studies of bulge dynamics using
a variety of probes.  Early work includes that of Minniti (1992) for K giants and bulge globular clusters, while recently, PNe have been employed (Beaulieau et al. 2000).  
A host of late-type stars have
been surveyed via radio techniques (e.g. SiC masers;
Izumiura et al. 1995).
With the completion of the 2MASS survey, I realized that M giants would make an ideal kinematic probe.  We sample from the red giant branch in $K$, $J-K$
without metallicity bias (see Rich et al. 2007).  
M giants have both Ti O bands and the Ca
infrared triplet, so there is certainty of obtaining excellent
radial velocities.  Mould (1983) measured the
first bulge velocity dispersion using M giants, while Sharples, Walker \& Cropper (1990) employed multiobject fiber spectroscopy to the problem, yielding
samples of $\sim 250$ stars.   M giants are advantageous
in that they represent a long lived evolutionary stage and are therefore common.  Further, they account for much of the light at 2-4$\mu \rm m$, the wavelengths for which maps of the bulge are constructed.  We find that the M giants are excellent probes, giving repeat measurements of $\sim 4\ km\ s^{-1}$.   

The Zhao (1996) self-consistent rapidly rotating bar model was fit to extent velocity data.  To best constrain the model, data spanning the
greatest range across the bulge are needed 
in $(l,b)$, along with a sample size large enough
to produce a credible line of sight distribution.   I concluded that
the {\sl BRAVA} survey would be the ideal path
for constraining the bulge model.  Examples of
the line of sight velocity distribution are found in 
Rich et al. (2007) and Figure 6 of this work.

Figure 3 shows our existing survey and our proposed
study for 2008.   Fields probing the edges of the bulge
will search for the cylindrical rotation that might be expected of a boxy pseudobulge (Kormendy \& Kennicutt 2004) and probe asymmetries predicted by the Zhao (1996) model.
We illustrate the resulting major axis rotation curve 
and l-v plot vs Zhao (1996) in Figure 4 compared with the Zhao rapidly rotating bar.  Note that our plots also include dynamics of a disk field at $(l,b)=-30^\circ,-4^\circ$.  We now have the benefit of hindsight: the K giant
fields of Minniti et al. (1992) appear to
fall on the {\sl BRAVA}rotation curve, but their
dispersion measurements lie below the M giants
(see Minniti \& Zoccali 2007), perhaps
indicative of disk contamination.  The bulge rotation 
curve
departs from a solid body at roughly $\pm 4 ^\circ $;
this is the first time such a departure is noted.

In my presentation, I asked how we would plot the BRAVA result on the
Binney (1978) diagram.  We estimate $\epsilon=1-e$
from Launhardt et al. (2002).  $V_{max}$ and $\sigma$ 
are from Figure 4.  We propose  $V_{max}= 75\ km\ s^{-1}$ and $sigma=115\ km\ s^{-1}$ (giving $0.65\pm 0.5$) with the main uncertainty
arising from our lack of an extragalactic perspective on the Milky Way.  One concern is whether there is a thick disk or bulge component outside of the central bulge isophotes that
is colder and and more rapidly rotating.  We place our bulge near that
of NGC 4565, sometimes proposed as a twin of the Milky Way.  The Galactic bulge {\it clearly}
falls below the oblate rotator line and near classical
bulges (as classified by Kormendy \& Kennicutt 2004) and is significantly, more slowly rotating than proposed pseudobulges.   As they note, it is not correct to classify
bars as "rotation supported" and this diagram is of limited utility when in fact, we have the Zhao model and a huge kinematic sample.

We have seen hints of substructure in our data; 6 fields clump at the 2.5$\sigma$ level.  Figure 6 shows that with an increased sample size, one candidate is not confirmed.  Substructure, arising from disrupting satellites, sub populations, or stars in unique orbit families, will be sought and confirmed by spatial coherence, kinematics, and abundance.   In the future, it would also be 
desirable to constrain the makeup of the bar in terms of age and metallicity.  Soto, Rich, \& Kuijken (2007) use proper motions and radial velocity to find some indication that the bar at $(l,b)=(0.9 ^\circ ,-4^\circ )$ is comprised of metal rich stars.


\end{document}